\documentclass[12pt,a4paper]{article}
\usepackage{latexsym}
\DeclareSymbolFont{AMSb}{U}{msb}{m}{n}
\DeclareSymbolFontAlphabet{\Bbb}{AMSb}
\newcommand{\N}{\Bbb{N}}
\newcommand{\Z}{\Bbb{Z}}
\newcommand{\R}{\Bbb{R}}
\newtheorem{theorem}{Theorem}
\newtheorem{lemma}{Lemma}

\newcommand{\tr}{\mathrm{tr} \ }
\newcommand{\proof}{\textbf{Proof.\ }}
\newcommand{\qed}{\begin{flushright} $\Box$ \end{flushright}}
\newcommand{\intdk}{\int_{\R^d}\frac{dk}{(2\pi)^d}}
\newcommand{\w}{\omega_\Lambda}

\begin{document}
\thispagestyle{empty}
\vspace*{-80pt} 
{\hspace*{\fill} Preprint-KUL-TF-97/32} 
\vspace{80pt} 
\begin{center} 
{\LARGE Non-extensive
  Bose-Einstein \\[7pt] Condensation Model}
 \\[25pt]  
 
{\large  
    T.Michoel\footnote{Aspirant van het Fonds voor Wetenschappelijk
    Onderzoek - Vlaanderen}\footnotetext{Email: {\tt tom.michoel@fys.kuleuven.ac.be}},
	A.Verbeure\footnote{Email: {\tt andre.verbeure@fys.kuleuven.ac.be}}
    } \\[75pt]   
{Instituut voor Theoretische Fysica} \\  
{Katholieke Universiteit Leuven} \\  
{Celestijnenlaan 200D} \\  
{B-3001 Leuven, Belgium}\\[40pt]
\end{center} 
\begin{abstract}\noindent
The imperfect Boson gas supplemented with a gentle repulsive interaction, is
completely solved. In particular it is proved that it has non-extensive
Bose-Einstein condensation, i.e. there is condensation without macroscopic
occupation of the ground ($k=0$) state level.\\[75pt]
\end{abstract}

\newpage
\section{Introduction}
The search for microscopic models of interacting Bosons showing Bose-Einstein
condensation is an ever challenging problem. It is known that the phenomenon
only appears for space dimensions $d\geq 3$ \cite{H}. A general two-body
interacting Bose system in a finite centered cubic box $\Lambda \subset \R^d$, with
volume $V=L^d$, is given by a Hamiltonian
\begin{equation}\label{gen-Ham}
H_\Lambda = T_\Lambda + U_\Lambda,
\end{equation}
where
\begin{eqnarray*}
T_\Lambda&=&\sum_{k\in\Lambda^*} \epsilon_k a^*_ka_k \; , \; \epsilon_k=
\frac{|k|^2}{2m}\\
U_\Lambda&=&\frac{1}{2V}\sum_{q,k,k'\in\Lambda^*}v(q)a^*_{k+q}a^*_{k'-q}a_{k'}
a_k\\
a(x)&=& \frac{1}{\sqrt{V}}\sum_{k\in\Lambda^*}a_k e^{ik.x}.
\end{eqnarray*}
The $a^\sharp(x)$ are the Boson operators satisfying the commutation rules
\begin{eqnarray*}
[a(x),a^*(y)]&=&\delta(x-y)\\
\lbrack a(x),a(y) \rbrack &=& 0,
\end{eqnarray*}
and
\[
\Lambda^*=\{k:k=\frac{2\pi}{L}n , n\in\Z^d\}.
\]
We limit ourself to periodic boundary conditions.

Rigorous results on the existence of Bose-Einstein condensation are known for
very special potentials $v$ in (\ref{gen-Ham}), in particular, of course for
$v=0$, the free Bose gas, and for $v$ in the $\delta$-function limit \cite{x}
or in the Van der Waals limit \cite{BLS}. Another class of models which are
treatable is this for which the Hamiltonian is a function of the number
operators $N_k=a^*_k a_k$ only. These models are called the diagonal models \cite{DLP}.
The Hamiltonian is a function of a set of mutually commuting operators with a
spectrum consisting of the integers. The operators can be considered as random
variables taking values in the integers. The equilibrium states are looked for
among the measures minimizing the free energy. This method, developed in a
series of papers \cite[and references therein]{DLP}, opened the possibility to
derive rigorous results for sofar unsolved interacting Bose gas models. The
method is a powerful application of the large deviation principle for quantum
systems.

In this paper we derive some rigorous results for another diagonal model,
inspired by \cite{S}, where the pressure is computed. We are not using the large
deviation technique of \cite{DLP}, but the full quantum mechanical technology,
in particular correlation inequalities, in order to prove the existence of
Bose-Einstein condensation. In section \ref{SM}, we first rederive the result of
\cite{S}, and give a concise, rigorous and direct proof of the pressure formula.
Some arguments of \cite{PZ} are translated into our situation. Our main
contribution is in section \ref{BEC}, where we proof the occurence of
Bose-Einstein condensation, and where we study in detail the type of
condensation. 

There exist different types of condensation. The best known is
macroscopic occupation of the ground state, but there is also so-called
generalized condensation, when the number of particles distributed over a set of
arbitrary small energies above the lowest energy level becomes macroscopic,
proportional to the volume. This notion has been put into a rigorous and
workable form in \cite{BLP}. 

As far as our results are concerned, this notion of generalized condensation is
crucial. We prove that in our model generalized condensation occurs without
macroscopic occupation of the ground state. As far as we know, this is the first
model of an interacting Bose gas for which this type of condensation is found.
The only existing result is for the free Bose gas, considering a special
thermodynamic limit, not of the type of increasing, absorbing cubes \cite{BLL,
BL}.

The result of section \ref{BEC} also allows, using the technique of \cite{FV},
to give an explicit form of the equilibrium states in the thermodynamic limit.
One verifies that they are of the same type as the equilibrium state of the
imperfect Bose gas.

\section{The model}\label{SM}
In \cite{S} Schr\"oder considers a Bose gas contained in a $d-$dimensional ($d\geq 3$) cubic box with Dirichlet
boundary conditions on two opposite faces and periodic boundary conditions on the remaining surface. This
can be interpreted as the model of a Bose gas enclosed between two hard walls at macroscopic distance. An
interaction term is introduced which behaves locally like the mean field interaction. This gives rise to
the following Hamiltonian:
\begin{equation}
H_\Lambda = \sum_{k \in \Lambda ^*}\epsilon_k N_{k,\Lambda} +
\frac{\lambda}{V}\left( N_{\Lambda}^2
+ \frac{1}{2}\sum_{j\in\N}\tilde N_{j,\Lambda}^2 \right),
\end{equation}
where
\begin{eqnarray*}
\Lambda &=& \{ x\in\R^d : -\frac{L}{2}\leq x_i \leq \frac{L}{2}, i=1,\ldots,
d-1; 0\leq x_d \leq L \}; V=L^d \\
\Lambda^* &=& \frac{2\pi}{L}\Z^{d-1} \times \frac{\pi}{L}\N \\
N_{k,\Lambda} &=& a^*(f_{k,\Lambda})a(f_{k,\Lambda}) \\
f_{k,\Lambda} &=& \left( \frac{2}{V} \right)^{1/2}\exp{[i(k_1x_1+ \cdots + k_{d-1}x_{d-1})]}\sin(k_dx_d)\\
\lambda &\in& \R^+ \\
\tilde N_{j,\Lambda} &=& \sum_{\{k\in\Lambda^*: k_d=\frac{\pi}{L}j\}}N_{k,\Lambda} \\
N_\Lambda &=&  \sum_{k \in \Lambda^*}N_{k,\Lambda}.\\
\end{eqnarray*}

Schr\"oder shows that the grand-canonical pressure of this so-called local mean field model coincides
with the grand-canonical pressure of the usual mean field model, or imperfect Bose gas, with Hamiltonian
\begin{equation}
H_\Lambda^{MF} = \sum_{k \in \Lambda^*}\epsilon_k N_{k,\Lambda} +
\frac{\lambda}{V}N_\Lambda^2,
\end{equation}
which is a soluble model.

From this result, Schr\"oder concludes that his model exhibits a phase transition with the same critical
behavior as the imperfect Bose gas, although macroscopic occupation of the ground state may not occur,
and opens the question whether generalized condensation, as defined in 
\cite{BLP}, does take place.

We study a model of an interacting Bose gas which is inspired by Schr\"oder's model,
but which contains a non-trivial part of the self-interaction terms appearing in the general
two-body repulsive interaction (\ref{gen-Ham}).  
More precisely, we
consider a system of identical bosons in a centered cubic box $\Lambda\in\R^d$, $d\geq3$, with volume $V=L^d$,
with periodic boundary conditions for the wave functions, and described by the Hamiltonian
\begin{equation}
H_\Lambda = \sum_{k \in \Lambda^*}\epsilon_k N_{k,\Lambda} + \frac{\lambda}{V}\left( N_\Lambda^2
+ \frac{1}{2}\sum_{k \in \Lambda^*} N_{k,\Lambda}^2 \right),
\end{equation}
where now
\begin{eqnarray*}
\Lambda^* &=& \frac{2\pi}{L}\Z^d  \\
N_{k,\Lambda} &=& a^*_{k,\Lambda} a_{k,\Lambda} \\
a^*_{k,\Lambda} &=& \frac{1}{\sqrt{V}}\int_{\Lambda}dx \ e^{ik.x} a^*(x) \\
\lambda &\in& \R^+ \\
N_\Lambda &=&  \sum_{k \in \Lambda^*}N_{k,\Lambda}.\\
\end{eqnarray*}

Our model can also be compared to the Huang-Yang-Luttinger model, rigorously 
studied in \cite{BLP2}. Compared to our model, here the interaction terms
$N_{k,\Lambda}^2$ appear with a minus sign and are therefore attractive 
perturbations of the imperfect Bose gas. The attractive character enhances 
(see \cite{BLP}) the condensation in the zero mode. The repulsive character 
of these terms in our model should make condensation in the zero mode more
difficult. Heuristically one might expect that such a model
is a candidate for non-extensive Bose-Einstein condensation.

First we give a new proof, inspired by a proof in \cite{PZ}, of the main result
of Schr\"oder, i.e. the equality of the grand-canonical pressure of this model and the grand-canonical
pressure of the imperfect Bose gas. From this we can immediately prove that there is no macroscopic
occupation of any single-particle state.

For every $\mu$ in $\R$, denote
\begin{equation}
H_\Lambda(\mu) = \sum_{k \in \Lambda^*}\epsilon_k N_{k,\Lambda} + \frac{\lambda}{V}\left( N_\Lambda^2
+ \frac{1}{2}\sum_{k \in \Lambda^*} N_{k,\Lambda}^2 \right) - \mu N_\Lambda,
\end{equation}
and
\begin{equation}\label{HMF}
H_\Lambda^{MF}(\mu) = \sum_{k \in \Lambda^*}\epsilon_k N_{k,\Lambda} + \frac{\lambda}{V}N_\Lambda^2
- \mu N_\Lambda.
\end{equation}
For $\alpha \leq 0$, let
\[
\mathcal{C}^\alpha = \{ t\in \mathcal{C}^b(\R^d): \inf_{k\in\R^d}(\epsilon_k - t_k - \alpha)>0\},
\]
with $\mathcal{C}^b(\R^d)$ the space of continuous bounded functions on $\R^d$. For
$t\in\mathcal{C}^\alpha$, let
\[
H_\Lambda^{t+\alpha} = \sum_{k \in \Lambda^*} (\epsilon_k - t_k - \alpha) N_{k,\Lambda}.
\]
First, we prove
\begin{lemma}\label{l1}
\begin{equation}\label{og}
\frac{1}{\beta V}\ln\tr e^{-\beta H_\Lambda(\mu)} \geq \frac{1}{\beta V}\ln\tr e^{-\beta H_\Lambda^{t+
\alpha}} - \frac{1}{V}\w^{t+\alpha}(H_\Lambda(\mu) - H_\Lambda^{t+\alpha}),
\end{equation}
with
\[
\w^{t+\alpha}(A) = \frac{\tr e^{-\beta H_\Lambda^{t+\alpha}}A}{\tr e^{-\beta
H_\Lambda^{t+\alpha}}}.
\]
\end{lemma}
\proof  The function $x\in [0,1]\mapsto \ln\tr e^{C+xD}$, for $C$ en $D$ self-adjoint is convex. Hence,
define the convex function $f$ on $[0,1]$ by
\begin{eqnarray*}
f(x) &=& \ln\tr e^{-\beta(x H_\Lambda(\mu) + (1-x)H_\Lambda^{t+\alpha})}.
\end{eqnarray*}
For all $a,b$ in $[0,1]$, $f(a)-f(b)-(a-b)f'(b)\geq 0$, in particular 
\[
f(1)\geq f(0)+f'(0),
\]
which immediately yields the stated inequality. 
\qed
We can now prove a first result.
\begin{theorem}\label{pres} 
The grand-canonical pressure at chemical potential $\mu$, 
\[
\tilde p(\mu)=\lim_{V\to\infty}\tilde p_\Lambda(\mu)=\lim_{V\to\infty}\frac{1}{\beta V}\ln\tr e^{-\beta
H_\Lambda(\mu)},
\]
exists for every $\mu$ in $\R$ and is given by
\[
\tilde p(\mu) = p^{MF}(\mu) = \inf_{\alpha\leq 0}\left( p(\alpha) + \frac{(\mu -
\alpha)^2}{4\lambda}\right),
\]
with $p^{MF}(\mu)$ the grand-canonical pressure of the imperfect Bose gas at chemical potential $\mu$ and
$p(\alpha)$ the free-gas grand-canonical pressure at chemical potential $\alpha$. 
\end{theorem}
(The expression for $p^{MF}(\mu)$ is computed in \cite{BLS}.)\\

\proof Since for every $\mu \in\R$, $H_\Lambda(\mu)\geq H_\Lambda^{MF}(\mu)$, we have
\[
\tilde p_\Lambda(\mu) \leq p^{MF}_\Lambda(\mu),
\]
and hence
\[
\limsup_{V\to\infty}\tilde p_\Lambda(\mu) \leq \lim_{V\to\infty}p^{MF}_\Lambda(\mu)=p^{MF}(\mu).
\]
To prove the lower bound, we make use of Lemma \ref{l1}. For $\alpha\leq 0$ and $t\in
\mathcal{C}^\alpha$, let
\[
\rho(k;t,\alpha)=\frac{1}{e^{\beta(\epsilon_k-t_k-\alpha)}-1}.
\]
Then
\begin{eqnarray*}
\w^{t+\alpha}(N_{k,\Lambda}) &=& \rho(k;t,\alpha) \\
\w^{t+\alpha}(N_{k,\Lambda} N_{k',\Lambda}) &=& \rho(k;t,\alpha)\rho(k';t,\alpha) 
\qquad \mathrm{if} \; k\not= k'\\
\w^{t+\alpha}(N_{k,\Lambda}^2) &=& \rho(k;t,\alpha)(2\rho(k;t,\alpha)+1).
\end{eqnarray*}
We calculate the r.h.s. of (\ref{og}). The first term gives
\[
\frac{1}{\beta V}\ln\tr e^{-\beta H_\Lambda^{t+\alpha}} = -\frac{1}{\beta V}\sum_{k\in\Lambda^*}
\ln\left( 1-e^{-\beta(\epsilon_k-t_k-\alpha)}\right).
\]
To calculate $\frac{1}{V}\w^{t+\alpha}(H_\Lambda(\mu))$, we write
\[
H_\Lambda(\mu) = \sum_{k \in \Lambda^*}(\epsilon_k - \mu) N_{k,\Lambda} +
\frac{\lambda}{V}
\sum_{k \in \Lambda^*}\sum_{k'\not=k \in \Lambda^*}N_{k,\Lambda}N_{k',\Lambda}
+ \frac{3\lambda}{2V}\sum_{k \in \Lambda^*} N_{k,\Lambda}^2,
\]
hence
\begin{eqnarray*}
\frac{1}{V}\w^{t+\alpha}(H_\Lambda(\mu))&=&\frac{1}{V}\sum_{k \in \Lambda^*}(\epsilon_k - \mu)
\rho(k;t,\alpha)\\&&+\frac{\lambda}{V^2}\sum_{k \in \Lambda^*}\sum_{k'\not=k \in \Lambda^*}
\rho(k;t,\alpha)\rho(k';t,\alpha) + \frac{c_V}{V},
\end{eqnarray*}
where
\[
c_V=\frac{3\lambda}{2V}\sum_{k \in \Lambda^*}\rho(k;t,\alpha)(2\rho(k;t,\alpha)+1).
\]
Also,
\[
\frac{1}{V}\w^{t+\alpha}(H_\Lambda^{t+\alpha}) =  \frac{1}{V}\sum_{k \in \Lambda^*}
(\epsilon_k - t_k - \alpha)\rho(k;t,\alpha).
\]
Substituting all this in (\ref{og}) we get
\begin{eqnarray*}
\tilde p_\Lambda(\mu) &\geq& -\frac{1}{\beta V}\sum_{k\in\Lambda^*}
\ln\left( 1-e^{-\beta(\epsilon_k-t_k-\alpha)}\right) \\
&&+ \frac{1}{V}\sum_{k \in
\Lambda^*}(\mu - t_k - \alpha)\rho(k;t,\alpha)\\
&&-\frac{\lambda}{V^2}\sum_{k \in \Lambda^*}\sum_{k'\not=k \in \Lambda^*}
\rho(k;t,\alpha)\rho(k';t,\alpha) - \frac{c_V}{V}.
\end{eqnarray*}
Since $\rho(k;t,\alpha)$ and $c_V$, for $V$ large enough, are bounded,
\begin{eqnarray}\label{linf}
\liminf_{V\to\infty}\tilde p_\Lambda(\mu) &\geq& 
-\beta^{-1} \intdk \ln\left( 1-e^{-\beta(\epsilon_k-t_k-\alpha)}\right) 
\nonumber \\
&&+\intdk (\mu - t_k - \alpha)
\rho(k;t,\alpha) \nonumber \\
&&-\lambda\left(\intdk \rho(k;t,\alpha) \right)^2.
\end{eqnarray}
For $\alpha\leq 0$ the free-gas pressure is given by
\[
p(\alpha)=-\beta^{-1} \intdk \ln\left( 1-e^{-\beta(\epsilon_k-\alpha)}\right),
\]
and
\[
p'(\alpha)=\intdk \frac{1}{e^{\beta(\epsilon_k-\alpha)}-1}.
\]
Also, let $\rho_c=p'(0)$ as usual.

First, consider the case $\mu< 2\lambda \rho_c$. Taking $\alpha<0$ and $t=0$ in
(\ref{linf}) we get
\begin{equation}\label{linf1}
\liminf_{V\to\infty}\tilde p_\Lambda(\mu) \geq p(\alpha) +
(\mu-\alpha)p'(\alpha)-\lambda(p'(\alpha))^2.
\end{equation}
For $\mu< 2\lambda \rho_c$, since $p'(\alpha)$ is increasing and $p'(0)=\rho_c$,
the equation
\[
p'(\alpha)=\frac{\mu-\alpha}{2\lambda}
\]
has a unique solution $\alpha^* <0$. Taking
$\alpha=\alpha^*$ in (\ref{linf1}) we get
\begin{eqnarray*}
\liminf_{V\to\infty}\tilde p_\Lambda(\mu) &\geq& p(\alpha^*)+
\frac{(\mu-\alpha^*)^2}{4\lambda} \\
&=& \inf_{\alpha\leq 0}\left\{ p(\alpha)+\frac{(\mu-\alpha)^2}{4\lambda}
\right\} \\
&=& p^{MF}(\mu),
\end{eqnarray*}
which proves the theorem for $\mu<2\lambda \rho_c$.

Consider now the case $\mu\geq 2\lambda \rho_c$. Take $\alpha=0$ and an
appropriate $t$ in (\ref{linf}):
\begin{eqnarray}\label{linf2}
\liminf_{V\to\infty}\tilde p_\Lambda(\mu) 
&\geq& -\beta^{-1}\intdk \ln\left(1-e^{-\beta(\epsilon_k-t_k)}\right) 
\nonumber\\
&&+\intdk (\mu - t_k)\rho(k;t) \nonumber \\
&&-\lambda\left( \intdk \rho(k;t) \right)^2,
\end{eqnarray}
with
\[
\rho(k;t)=\frac{1}{e^{\beta(\epsilon_k-t_k)}-1}.
\]
For all $\delta >0,$ take $t_\delta \in \mathcal{C}^0$ such that
\[
t_\delta(k)=0, \; |k|>\delta.
\]
Then
\[
\intdk \rho(k;t_\delta)=\int_{|k|\leq\delta}\frac{dk}{(2\pi)^d} \rho(k;t_\delta)
+ \int_{|k|>\delta}\frac{dk}{(2\pi)^d}\frac{1}{e^{\beta\epsilon_k}-1}.
\]
Letting $\delta\to 0,$ the second term in the r.h.s. converges to $\rho_c$. Take
$t_\delta$ such that the first term in the r.h.s. converges to
$\frac{\mu}{2\lambda}-\rho_c$ as $\delta\to 0$. Such a sequence of $t_\delta$'s
can be constructed rigorously by using the Approximation theorem proved
in \cite{BDLP}.
It certainly means that $t_\delta\to 0$ as $\delta\to 0$. Hence we get
\[
\intdk \rho(k;t_\delta)\to\frac{\mu}{2\lambda}
\]
as $\delta\to 0$, and thus
\begin{eqnarray*}
\liminf_{V\to\infty}\tilde p_\Lambda(\mu) &\geq& p(0) + \frac{\mu^2}{4\lambda}\\
&\geq& \inf_{\alpha\leq 0}\left\{ p(\alpha)+\frac{(\mu-\alpha)^2}{4\lambda}
\right\} \\
&=& p^{MF}(\mu),
\end{eqnarray*}
so that the theorem is proved for $\mu\geq 2\lambda\rho_c$ as well.
\qed
From Theorem \ref{pres} we can immediately derive that there is no macroscopic
occupation of any single-particle state, in particular:
\begin{theorem}\label{p1}
For every $\epsilon>0$ and for $V$ large enough, we have for every $k\in\Lambda^*$:
\[
\frac{1}{V}\omega_\Lambda(N_{k,\Lambda})<\epsilon,
\]
where $\omega_\Lambda$ is the finite-volume Gibbs state of $H_\Lambda(\mu)$.
\end{theorem}
\proof We have
\begin{eqnarray*}
e^{\beta Vp^{MF}_\Lambda(\mu)} 
&=& \tr e^{-\beta H_\Lambda^{MF}(\mu)}\\
&=& \tr \left( e^{-\beta H_\Lambda(\mu)}e^{\frac{\beta\lambda}{2V}\sum_{k\in
\Lambda^*}N_{k,\Lambda}^2} \right)\\
&=& \w\left( e^{\frac{\beta\lambda}{2V}\sum_{k\in
\Lambda^*}N_{k,\Lambda}^2} \right) e^{\beta V \tilde p_\Lambda(\mu)}.
\end{eqnarray*}
Hence,
\[
p^{MF}_\Lambda = \tilde p_\Lambda(\mu) + \frac{1}{\beta V}\ln
\w\left( e^{\frac{\beta\lambda}{2V}\sum_{k\in
\Lambda^*}N_{k,\Lambda}^2} \right).
\]
By Theorem \ref{pres} we get:
\[
\lim_{V\to\infty}\frac{1}{\beta V}\ln\w\left( e^{\frac{\beta\lambda}
{2V}\sum_{k\in\Lambda^*}N_{k,\Lambda}^2} \right) = 0.
\]
From the Jensen inequality, i.e. for $F$ a convex function and $\omega$ a normal
state
\[
\omega(F(X))\geq F(\omega(X)),
\]
we get
\[
\w\left( e^{\frac{\beta\lambda}{2V}\sum_{k\in\Lambda^*}N_{k,\Lambda}^2}
\right) \geq e^{\frac{\beta\lambda}{2V}\sum_{k\in\Lambda^*}\w(N_{k,\Lambda}^2)},
\]
or
\[
0\leq \frac{\lambda}{2V^2}\sum_{k\in\Lambda^*}\w(N_{k,\Lambda}^2) \leq
\frac{1}{\beta V}\ln\w\left( e^{\frac{\beta\lambda}
{2V}\sum_{k\in\Lambda^*}N_{k,\Lambda}^2} \right).
\]
Hence
\[
\lim_{V\to\infty}\frac{1}{V^2}\sum_{k\in\Lambda^*}\w(N_{k,\Lambda}^2)=0.
\]
Since for each $k\in\Lambda^*$
\[
0\leq\left( \frac{1}{V}\w(N_{k,\Lambda}) \right)^2\leq \frac{1}{V^2}
\w(N_{k,\Lambda}^2)\leq \frac{1}{V^2}\sum_{k'\in\Lambda^*}\w(N_{k',\Lambda}^2),
\]
we get the Theorem.
\qed

\section{Bose-Einstein condensation}\label{BEC}
In \cite{BLP} it is stressed that Bose condensation does not necessarily manifest
itself through a macroscopic occupation of a single-particle state (the ground
state usually), but that there are in fact two good candidates for the concept
of macroscopic occupation of the zero-kinetic energy state. Macroscopic
occupation of the ground state is said to occur when the number of particles in
the ground state becomes proportional to the volume; generalized condensation is
said to occur when the number of particles whose energy levels lie in an
arbitrary small band above zero becomes proportional to the volume. Obviously,
the first implies the second. However, the  second can occur without the first;
this is called non-extensive condensation. The concept of generalized
condensation was first introduced in \cite{G}.

More precisely, we have:
\begin{enumerate}
\item macroscopic occupation of the ground state if the limit
\[
\lim_{V\to\infty}\frac{1}{V}\w(N_{0,\Lambda})
\]
exists and is strictly positive,

\item generalized condensation if the limit
\[
\lim_{\delta\to0}\lim_{V\to\infty}\frac{1}{V}\sum_{\{k\in\Lambda^*:
\epsilon_k < \delta\} }\w(N_{k,\Lambda})
\]
exists and is strictly positive,

\item non-extensive condensation if the limit
\[
\lim_{V\to\infty}\frac{1}{V}\w(N_{0,\Lambda})=0,
\]
but nevertheless the limit
\[
\lim_{\delta\to0}\lim_{V\to\infty}\frac{1}{V}\sum_{\{k\in\Lambda^*:
\epsilon_k < \delta\} }\w(N_{k,\Lambda})
\]
exists and is strictly positive.
\end{enumerate}
Examples of these different occurences of Bose condensation in the free Bose
gas, depending on how the bulk limit is taken, can be found in \cite{BLP,BLL,BL}.

As is proved in Theorem \ref{p1}, there is no macroscopic occupation of the
ground state in our system. However, as we will show, there is generalized
condensation. In other words, we have a model for an interacting Bose gas
displaying non-extensive condensation.

Our approach is based on \cite{FV}, where the imperfect Bose gas is treated.
The system is given by the local Hamiltonian $H_\Lambda$, with periodic
boundary conditions
\begin{equation}\label{H}
H_\Lambda = \sum_{k \in \Lambda^*}\epsilon_k N_{k,\Lambda} +
\frac{\lambda}{V}\left( N_\Lambda^2
+ \frac{1}{2}\sum_{k \in \Lambda^*} N_{k,\Lambda}^2 \right) - \mu_\Lambda
N_\Lambda,
\end{equation}
as specified before, and $\mu_\Lambda$ is determined by the constant density
$\rho>0$ equation:
\[
\frac{1}{V}\w(N_\Lambda)=\rho.
\]

We study the equilibrium state of this system in the grand-canonical
ensemble. The key technique is the equivalence of the equilibrium condition
or Gibbs state $\omega_\Lambda$ with the correlation inequalities \cite{FV2,FV3}
\begin{equation}\label{cor}
\beta\w\left(X^*[H_\Lambda,X]\right) \geq \w(X^*X)\ln\frac{\w(X^*X)}{\w(XX^*)},
\end{equation}
for all local observables $X$ belonging to the domain of 
$[H_\Lambda,.]$. In particular, we take for $X$ polynomials in the creation and
annihilation operators.
We prove the occurence of non-extensive condensation in this model, and follow
closely the method used in \cite{FV}.

\begin{lemma}\label{l3}
$\forall k,j\in\Lambda^*$:
\begin{enumerate}
\item \[
\beta\w\left(-\epsilon_kN_{k,\Lambda} +(\mu_\Lambda-\frac{2\lambda}{V}
N_\Lambda)N_{k,\Lambda} - \frac{\lambda}{V}N_{k,\Lambda}^2
+\frac{3\lambda}{2V}N_{k,\Lambda}\right)
\]
\begin{equation}\label{in1}
\geq \w(N_{k,\Lambda})\ln\frac{\w(N_{k,\Lambda})}
{\w(N_{k,\Lambda})+1}\; ;
\end{equation}

\item \begin{equation}\label{in2}
\w\left( (\mu_\Lambda-\frac{2\lambda}{V}N_\Lambda)N_{k,\Lambda}\right)
\leq \w\left(\epsilon_j N_{k,\Lambda} + \frac{4\lambda}{V}N_{j,\Lambda}
N_{k,\Lambda} + \frac{3\lambda}{2V} N_{k,\Lambda} \right).
\end{equation}
\end{enumerate}
\end{lemma}
\proof For \emph{(i)}, the result follows by taking $X=a_k$ in the correlation
inequality (\ref{cor}). One gets \emph{(ii)} by taking
\[
X=a_jN_{k,\Lambda}^{1/2},
\]
in the inequality
\[
\w([X^*,[H_\Lambda,X]])\geq 0,
\]
which follows immediately from (\ref{cor}) by adding the correlation 
inequality for 
$X$ and the complex conjugate of the correlation inequality for 
$X^*$. 
\qed
\begin{lemma}\label{l4}
For every $\delta >0$, for every $V$  and for every  
$k\in \Lambda^*,|k|\geq\delta$,  
\[
\w(N_{k,\Lambda})\leq\frac{1}{e^{c_k(\Lambda)}-1}+
\frac{4\lambda}{V} \w(N_{j,\Lambda}N_{k,\Lambda})\frac{1}
 {1-e^{-c_\delta(\Lambda)}},
\]
with
\[
c_k(\Lambda) = \beta\left( \epsilon_k -\frac{\delta^2}{8m}- \frac{3\lambda}{V}, 
\right),
\]
$c_\delta(\Lambda)=c_k(\Lambda)|_{|k|=\delta}$
and $j\in \Lambda^*,|j|\leq\frac{\delta}{2}$.
\end{lemma}
\proof Substitution of (\ref{in2}) in (\ref{in1}),  changing the sign, and using
the trivial bound $\w(N_{k,\Lambda}^2)\geq 0$ we get
\[
\beta(\epsilon_k-\epsilon_j-\frac{3\lambda}{V})\w(N_{k,\Lambda}) - \frac{4
\lambda}{V} \w(N_{j,\Lambda}N_{k,\Lambda})
\]
\begin{equation}\label{in3}
\leq \w(N_{k,\Lambda})\ln\frac{\w(N_{k,\Lambda})+1}
{\w(N_{k,\Lambda})}. 
\end{equation}

Take $\delta >0$ arbitrary,  $|k|\geq\delta$  and $|j|\leq \frac{\delta}{2}$.

Using $\epsilon_j\leq\frac{\delta^2}{8m}$, (\ref{in3}) becomes
\[
c_k(\Lambda)\w(N_{k,\Lambda})-\frac{4
\lambda}{V} \w(N_{j,\Lambda}N_{k,\Lambda})\leq\w(N_{k,\Lambda}) 
\ln\frac{\w(N_{k,\Lambda})+1}{\w(N_{k,\Lambda})}.
\]
The Lemma now follows from convexity arguments in the r.h.s.: we want to solve
for $t$ the inequality
\[
ct-b\leq t\ln\frac{t+1}{t},
\]
with $c$ and $b$ positive constants and $t\in\R^+$. It follows that
$t\leq t_2$, with $t_2$ satisfying
\[
ct_2-b = t_2\ln\frac{t_2+1}{t_2}.
\]
One can write this as $t\leq t_1+(t_2-t_1)$, with 
\[
t_1=\frac{1}{e^c-1}.
\]
Let $f(t)=t\ln\frac{t+1}{t}$, $f$ is concave, hence
\[
f(t_2)-f(t_1)-(t_2-t_1)f'(t_1)\leq 0,
\]
and
\[
t_2-t_1\leq b\frac{1}{1-e^{-c}}.
\]
Substitute this into the inequality $t\leq t_1+(t_2-t_1)$, one gets
\[
t\leq\frac{1}{e^c-1}+b\frac{1}{1-e^{-c}}.
\]
Finally, use $|k|\geq\delta$ in the second term on the r.h.s. to prove the Lemma.
\qed

\begin{lemma}\label{l5}
For every $\epsilon>0$,  $V$ large enough and $j\in\Lambda^*$:
\[
\frac{1}{V^2}\w(N_\Lambda N_{j,\Lambda}) < \epsilon.
\]
\end{lemma}
\proof
(\ref{in1}) gives
\[
\beta\w\left(-\epsilon_jN_{j,\Lambda} +(\mu_\Lambda-\frac{2\lambda}{V}
N_\Lambda)N_{j,\Lambda}
+\frac{3\lambda}{2V}N_{j,\Lambda}\right)
\]
\[
\geq \w(N_{j,\Lambda})\ln\frac{\w(N_{j,\Lambda})}
{\w(N_{j,\Lambda})+1}\geq -1.
\]
This can be rewritten in the form
\begin{equation}\label{aap}
\frac{2\lambda}{V^2}\w(N_\Lambda N_{j,\Lambda})\leq \frac{1}{\beta V}+(\mu_\Lambda
+\frac{3\lambda}{2V}-\epsilon_j)\frac{1}{V}\w(N_{j,\Lambda}).
\end{equation}
Taking $X=a_j$ in the inequality 
\[
\w([X^*,[H_\Lambda,X]])\geq 0,
\]
gives
\[
\mu_\Lambda\leq 2 \lambda \rho + \epsilon_j + \frac{4\lambda}{V}\w(N_{j,\Lambda})+
\frac{3\lambda}{2V}.
\]
Putting this into (\ref{aap}) gives
\[
\frac{2\lambda}{V^2}\w(N_\Lambda N_{j,\Lambda})\leq \frac{1}{\beta V} + 
(2 \lambda \rho + \frac{3\lambda}{V})\frac{1}{V}\w(N_{j,\Lambda}) +
\frac{4\lambda}{V^2}\w(N_{j,\Lambda})^2
\]
Using Theorem \ref{p1}  proves the Lemma.
\qed

We now prove the existence of generalized condensation in the thermodynamic
limit $V\to\infty$, taken with constant particle density $\rho$.
\begin{theorem}\label{t2}
One has
\begin{enumerate}
\item 
\[
\lim_{\delta\to 0}\lim_{V\to\infty}\frac{1}{V}\sum_{\{k\in\Lambda^*:
|k|<\delta\}}\w(N_{k,\Lambda})\geq \rho - \intdk\frac{1}{e^{\beta\epsilon_k}-1};
\]

\item for every $\rho >0$, there is a $\beta_c$ such that for all $\beta >
\beta_c$:
\[
0<\lim_{\delta\to 0}\liminf_{V\to\infty}\frac{1}{V}\sum_{\{k\in\Lambda^*:
|k|<\delta\}}\w(N_{k,\Lambda})
\]
\[
\leq \lim_{\delta\to
0}\limsup_{V\to\infty}\frac{1}{V}\sum_{\{k\in\Lambda^*:
|k|<\delta\}}\w(N_{k,\Lambda})\leq \rho.
\]
\end{enumerate}
\end{theorem}
\proof We have clearly
\[
\frac{1}{V}\sum_{\{k\in\Lambda^*:|k|<\delta\}}\w(N_{k,\Lambda}) = \rho -
\frac{1}{V}\sum_{\{k\in\Lambda^*:|k|\geq\delta\}}\w(N_{k,\Lambda}).
\]
Applying Lemma \ref{l4} gives
\[
\frac{1}{V}\sum_{\{k\in\Lambda^*:|k|<\delta\}}\w(N_{k,\Lambda})
\geq \rho - \frac{1}{V}\sum_{\{k\in\Lambda^*:|k|\geq\delta\}}
\frac{1}{e^{c_k(\Lambda)}-1}
\]
\begin{equation}\label{equ15}
-\frac{4\lambda}{V^2}\sum_{\{k\in\Lambda^*:|k|\geq\delta\}}
 \w(N_{j,\Lambda}N_{k,\Lambda})\frac{1}
 {1-e^{-c_\delta(\Lambda)}}.
\end{equation}

Take $\epsilon >0$ arbitrary, and $V$ large enough such that Lemma \ref{l5} is satisfied.
This implies  that
\[
\frac{1}{V^2}\sum_{\{k\in\Lambda^*:|k|\geq\delta\}} \w(N_{j,\Lambda}N_{k,\Lambda})
\leq \frac{1}{V^2}\w(N_\Lambda N_{j,\Lambda}) < \epsilon.
\]
Hence taking $V$ large enough, the second term in the r.h.s. of (\ref{equ15}) can be made 
arbitrarily close to
\[
\int_{|k|\geq\delta}\frac{dk}{(2\pi)^d}\frac{1}{e^{\beta(\epsilon_k-\frac{\delta^2}
{8m})}-1},
\] 
whereas the third term is made arbitrarily small.

Hence in the limit $V\to\infty$, one gets
\[
\lim_{V\to\infty}\frac{1}{V}\sum_{\{k\in\Lambda^*:|k|<\delta\}}\w(N_{k,\Lambda})
\geq \rho - \int_{|k|\geq\delta}\frac{dk}{(2\pi)^d}\frac{1}{e^{\beta(\epsilon_k-\frac{\delta^2}
{8m})}-1}.
\]
Now take the limit $\delta\to 0$ to get \emph{(i)}.

The function 
\[
\beta\mapsto f(\beta)=\intdk\frac{1}{e^{\beta\epsilon_k}-1}
\]
is clearly decreasing in $\beta >0$ and furthermore $f(\beta)\to\infty$ for
$\beta\to 0$, and $f(\beta)\to 0$ for $\beta\to\infty$. Hence for every $\rho>0$
there exists $\beta_c >0$, defined by
\[
\rho=\intdk\frac{1}{e^{\beta_c\epsilon_k}-1}.
\]
Together with \emph{(i)} this proves \emph{(ii)}.
\qed

Theorem \ref{pres} proves that the model (\ref{H}) has the same pressure as the
imperfect Bose gas. Theorems \ref{p1} and \ref{t2} prove that the model (\ref{H}) shows a
Bose-Einstein condensation exactly as the imperfect Bose gas, be it that the
nature of the condensation is different. One aspect of this is that the ground
state ($k=0$) condensation of the imperfect Bose gas is unstable against any
arbitrary small repulsive perturbation of the type $\frac{\gamma}{V}\sum
_{k\in\Lambda^*}N_{k,\Lambda}^2$, for any $\gamma >0$. The condensation becomes
non-extensive. However on the level of the thermodynamics the models are
similar.

The natural question to ask is, whether the equilibrium states of the two models
coincide. For the imperfect Bose gas, this problem is solved e.g. in \cite{FV}.
We are not going into the details, but the technique of \cite{FV} can also be
used in order to solve rigorously the equilibrium- or KMS-equations of our
model. The result is that all equilibrium states are of the same type as the
ones of the imperfect Bose gas. In particular, the equilibrium states are also
integrals over a set of quasi-free or generalized free states.

On the other hand it is interesting to remark the following. Given this result,
one might ask whether the variational principle of statistical mechanics,
formulated in the thermodynamic limit, but restricted to the set of quasi-free
states, does also give the results of this paper, namely the existence of
condensation and the equilibrium states. Performing this program, one remarks
that the particular type of condensation is not recovered by this method. Hence
for the time being, the only way to keep track of it, is to follow closely the
details of the thermodynamic limit as is done above. This work illustrates
clearly that care must be taken of this limit and that statistical mechanics
remains the theory of really handling the thermodynamic limit.

\section*{Acknowledgements}
The authors thank the referee for his careful reading of the paper, leading to a 
better version of it.

\end{document}